\documentclass[twocolumn,superscriptaddress,pra]{revtex4-1}
\usepackage{color}
\usepackage{xspace}
\usepackage{amsmath}
\usepackage{amssymb}
\usepackage{amsbsy}
\usepackage{graphicx}
 \usepackage{bm}
 \usepackage{float}
\usepackage{relsize}
\usepackage[normalem]{ulem}

\begin{document}

\title{Static-response theory and the roton-maxon spectrum of a flattened dipolar Bose-Einstein condensate}

\author{R.~N.~Bisset}
\email{russell.bisset@itp.uni-hannover.de}
\affiliation{INO--CNR BEC Center and Dipartimento di Fisica, Universit\`a di Trento, 38123 Povo, Italy}
\affiliation{Institut f\"ur Theoretische Physik, Leibniz Universit\"at Hannover, 30167 Hannover, Germany}

\author{P.~B.~Blakie}
\affiliation{Department of Physics, Centre for Quantum Science,
and Dodd-Walls Centre for Photonic and Quantum Technologies, University of Otago, Dunedin 9016, New Zealand}

\author{S.~Stringari}
\affiliation{INO--CNR BEC Center and Dipartimento di Fisica, Universit\`a di Trento, 38123 Povo, Italy}

\begin{abstract}

Important information for the roton-maxon spectrum of a flattened  dipolar Bose-Einstein condensate is extracted by applying a static perturbation exhibiting a periodic in-plane modulation.
By solving the Gross-Pitaevskii equation in the presence of the weak perturbation we evaluate the linear density response of the system and use it, together with sum rules, to provide a Feynman-like upper-bound prediction for the excitation spectrum, finding excellent agreement with the predictions of full Bogoliubov calculations. By suddenly removing the static perturbation, while still maintaining the trap, we find that the density modulations -- as well as the weights of the perturbation-induced side peaks of the momentum distribution -- undergo an oscillatory behavior with double the characteristic frequency of the excitation spectrum.
The measurement of the oscillation periods could provide an easy determination of dispersion relations.

\end{abstract}

\maketitle


\section{Introduction}

The quasiparticle energy dispersion $\epsilon(k)$ -- for momentum $k$ -- directly underpins the correlations and fluctuations of quantum fluids. 
An intriguing example is the excitation spectrum of superfluid $^4$He which exhibits a characteristic local  minimum in a roton region \cite{Landau1947,Feynman1954}.

Dilute quantum gases offer many parallels with dense quantum liquids in highly-controllable settings.
An exemplary system is the dipolar Bose-Einstein condensate (BEC), now producible with highly-magnetic atoms of chromium \cite{Griesmaier2005a,Beaufils2008}, dysprosium \cite{Mingwu2011a,Kadau2016a} or erbium \cite{Aikawa2012a}.
While remaining in the weakly-interacting regime, these systems possess several phenomena reminiscent of superfluid $^4$He thanks to the long-ranged and anisotropic nature of dipole-dipole interactions \cite{Lahaye_RepProgPhys_2009,Baranov2012,Pitaevskii16}.
A remarkable example is the recent production of dilute self-bound droplets \cite{Kadau2016a,Chomaz2016,Schmitt2016}, having liquid properties, and stabilized by quantum fluctuations \cite{Wachtler2016b,Baillie2016}\footnote{Note that related self-bound droplets were also observed in binary BECs \cite{Petrov2015a,Cabrera2018,Semeghini2018}}.
Another parallel is the prediction of a  supersolid phase \cite{Baillie2018,Roccuzzo2018,Saito2009}, whose experimental realization has been the subject of recent significant advances \cite{Tanzi2018,Bottcher2019,Chomaz2019}.

An important parallel with superfluid $^4$He concerns the roton-maxon dispersion.
While the rotons of $^4$He rely on strong correlations, it is remarkable that an analogous dispersion was predicted in 2003 to occur for weakly-interacting dipolar condensates \cite{Santos2003a,Giovanazzi2004}.
Over the last year, landmark experiments have produced the first evidence for dipolar rotons \cite{Chomaz2018}, as well as the first glimpses of the roton-maxon spectrum using Bragg spectroscopy \cite{Petter2018} (see related theory \cite{Blakie2012a}).
There has also been intense interest in rotons of other weakly-interacting BECs such as with shaken optical lattices \cite{Ha2015}, synthetic spin-orbit coupling \cite{Martone2012,Khamehchi2014,Ji2015}, and in the presence of a cavity \cite{Leonard2017}. Dipolar rotons are fundamentally different, though, since they genuinely arise from interactions and are not induced by external driving.

An important finding of the Bragg-spectroscopy experiment \cite{Petter2018} was the confirmation that the roton energy rapidly vanishes as instability is approached. Crucially, though, the authors of \cite{Petter2018} found significant deviations from the predictions of the prevailing theory, which includes quantum fluctuations in a local-density approximtion. Such  an approach underpins ongoing studies of self-bound droplets and dipolar supersolids, and measurements of the roton-maxon spectrum can furnish a highly-sensitive test for the development of improved theoretical descriptions.

Among the key challenges for measuring the dipolar roton-maxon spectrum is the requirement for the condensate to be highly anisotropic, with a short axis along the direction of dipole polarization.
The existence of rotons also creates a vulnerability to condensate collapse \cite{Chomaz2018}, that can even be triggered by thermal density fluctuations \cite{Linscott2014a}.
Previous proposals to detect rotons were based on applying a 1D lattice to either trigger a roton collapse of the condensate \cite{Corson2013a,Corson2013b}, or to detect a peak of the momentum distribution for lattice wavelengths near the roton minimum \cite{JonaLasinio2013b}, but these did not consider how to extract the dispersion relation itself.

We develop novel approaches to extract the roton-maxon spectrum based on the application of a static 1D lattice in the plane of a flattened dipolar BEC.
To demonstrate their utility we focus on the radially unconfined geometry, with a harmonic trap only along the direction of dipole polarization.
The response of the density is highly sensitive to the lattice wavelength and, with the help of sum rules, can be used to provide a rigorous upper bound for the energy dispersion.
We calculate this upper bound numerically, using a 3D Gross-Pitaevskii equation (GPE), and compare it with the exact prediction for the roton-maxon dispersion obtained directly from Bogoliubov-de Gennes (BdG) calculations, finding an almost exact agreement.
To compliment the possibility of extracting the density response in position space using \emph{in situ} imaging, we demonstrate that the side peaks of the momentum distribution - of relevance to expansion experiments - can also be used to give the dispersion relation.
Finally, we show that if the static lattice is suddenly removed, while the trap remains on, the system exhibits an oscillatory behavior in position space, as well as for the momentum side peaks, which  provides a means to extract the dispersion relation of the excitation spectrum  without having to calibrate the lattice strength or the magnitude of the density response.
Intriguingly, we find a phase inversion of the momentum side peak oscillations for rotons compared to maxons, which is quantitatively described by our perturbation theory without any fitting parameters.

\begin{figure}
\begin{center}
\includegraphics[width=3.4in]{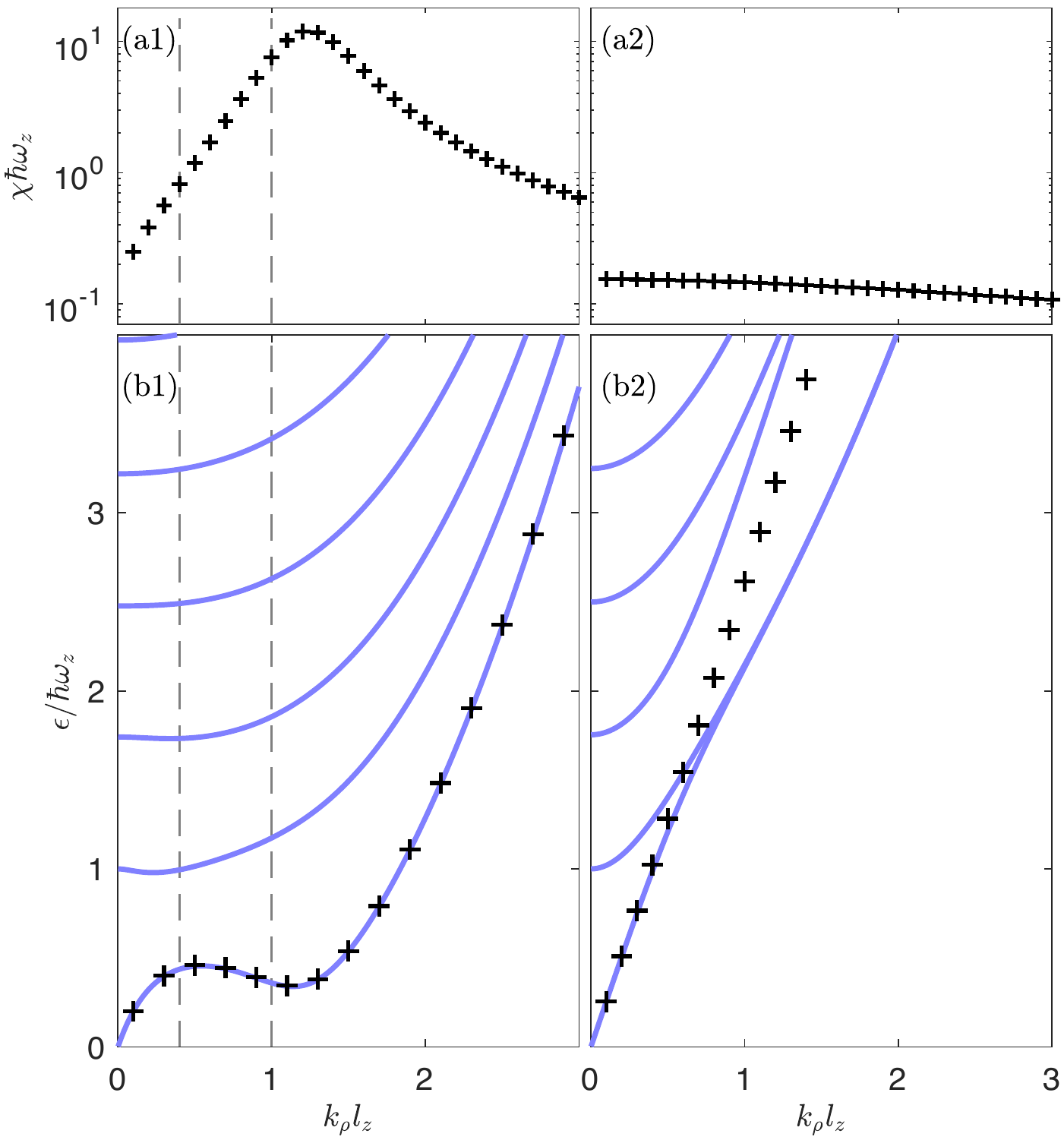}
\caption{a) GPE prediction for the static density response [Eq.~(\ref{Eq:StatResPos})] versus in-plane momentum, and
(b) energy dispersion for (a1,b1) a dipolar and (a2,b2) a non-dipolar condensate. The BdG energy bands appear as blue solid lines, while the sum-rule upper bound -- based on the the density response of the GPE ground state to a perturbing lattice potential with $k_\rho = |\mathbf{k}_{\rm L}|$ (\ref{Eq:StatResPos})-(\ref{Eq:EUpBound}) -- is marked with plus symbols.
(b1) The sum rule's ability to provide an almost exact prediction for the roton-maxon spectrum of the lowest band is highly nontrivial given our 3D GPE calculations inherently account for contributions from higher bands. (b2)
A similar sum-rule prediction for a non-dipolar BEC exhibits good agreement with the lowest BdG band only for $k_\rho l_z\lesssim 1$.
Both regimes have $\mu=9.5\hbar\omega_z$ and compressibility $\hbar\omega_z\chi(k_\rho\to0)=0.15$, and for the static response calculations we use $V_{\rm L}=0.0025\hbar\omega_z$.
Vertical dashed lines represent the cases in Figs.~\ref{Fig:Dyn_NL} and \ref{Fig:E_V0}.
\label{Fig:Dispersion}}
\end{center}
\end{figure}

\section{Formalism}

We consider a 3D flattened dipolar BEC that is harmonically trapped only along the $z$ direction, characterized by frequency $\omega_z$. Along the untrapped directions the components of the in-plane wavevector $\mathbf{k}_\rho=(k_x,k_y)$ provide  good quantum numbers. No assumptions are made about the density profile along the $z$ direction and this must be solved numerically. With regard to this last point, it was demonstrated that accurate treatment of the tight direction can be crucial for providing qualitatively useful results \cite{Baillie2015a}.

The primary motivation for considering the radially untrapped regime is that in the presence of harmonic trapping rotons are strongly `attracted' to high density, tightly confining them to a small central region \cite{JonaLasinio2013b,Bisset2013a} and  reducing the rotonized portion of the system and the corresponding observable signal \cite{JonaLasinio2013}.
Nevertheless, as a check, we have also performed calculations in the presence of harmonic trapping in all directions (not shown here), and observe qualitatively consistent results, with the main difference being that each excitation exhibits a momentum broadening.

The generalized GPE takes the form \cite{Wachtler2016a,Lima2011a,Bisset2016,FerrierBarbut2016,Chomaz2016}
\begin{align}
i\hbar & \frac{\partial\psi(\mathbf{x})}{\partial t} = \Big[-\frac{\hbar^2\nabla^2}{2m} + \frac{m\omega_z^2z^2}{2} \label{Eq:GPE} \\
&+ \int d^3\mathbf{x}^\prime U(\mathbf{x}-\mathbf{x}^\prime)|\psi(\mathbf{x}^\prime)|^2
+ \gamma_{\rm QF}|\psi|^3 \Big] \psi(\mathbf{x}) , \notag 
\end{align}
with the interaction potential being well-described by the pseudopotential
$U(\mathbf{r})=g_s\delta(\mathbf{r}) + U_{\rm dd}(\mathbf{r})$. The contact interaction strength is $g_s = 4\pi a_s\hbar^2/m$, for s-wave scattering length $a_s$ and mass $m$.
The dipoles are polarized along $z$ and the corresponding dipole-dipole interactions are described by
$U_{\rm dd}(\mathbf{r}) = (3g_{\rm dd}/4\pi)(1-3\cos^2\theta)/r^3$, where $\theta$ is the angle between $\mathbf{r}$ and the $z$ axis. Their strength is given by
$g_{\rm dd} = \mu_0\mu_m^2/3$,
for magnetic dipole moment $\mu_m$
\footnote{To prevent Fourier copies along the $z$ direction from interacting we truncate the range of the dipole-dipole interaction \cite{Ronen2006a}.}.
The dipolar Lee-Huang-Yang (LHY) correction is added in the local density sense, being proportional to
$\gamma_{\rm QF} =(32g_{\rm s}/3)\sqrt{a_{\rm s}^3/\pi}(1+3\epsilon_{\rm dd}^2/2)$ \cite{Lima2011a,Lima2012a},
where the ratio $\epsilon_{\rm dd} = g_{\rm dd}/g_s$ is useful since $\epsilon_{\rm dd}>1$ signals the dipole-dominated regime.
It should be noted that the main effect of the LHY term throughout this paper is to shift the scattering length of the roton instability downwards by around 8\%. The results otherwise remain qualitatively the same.

To benchmark our approach we obtain excitation energies and wavefunctions by solving the BdG equations.
These can be obtained by linearizing about Eq.~(\ref{Eq:GPE}) in the absence of any perturbing lattice \cite{Baillie2017}. Solving these in the present regime cannot be done analytically, so we use the numerical techniques outlined in \cite{Baillie2015a} but here we include the LHY term.

\section{Sum rules and the static density response}

We consider the condensate response to the 1D periodic lattice perturbation
\begin{equation}
V_{\rm pert} = 2V_{\rm L} \cos (\mathbf{k}_{\rm L}\cdot \mathbf{x}), \label{Eq:VLatt}
\end{equation}
where $V_{\rm L}$ is a constant and $\mathbf{k}_{\rm L} = (k_{\rm L},0,0)$.
To do this we solve for ground states of the time-independent GPE including $V_{\rm pert}$.
In the limit of small $V_{\rm L}$
the spatial density oscillation arising from the perturbation
furnishes the static density response function
\begin{equation}
\chi(\mathbf{k}_{\rm L}) = \lim_{V_{\rm L}\to0}\frac{\Delta n}{2V_{\rm L}} , \label{Eq:StatResPos}
\end{equation}
where the amplitude of the density perturbation is
\begin{equation}
\Delta n = \frac{\max\{n(x)\}-\min\{n(x)\}}{2n_0}, \label{Eq:Delta_n}
\end{equation}
for the 2D density $n(x) = \int |\psi(\mathbf{x})|^2 dz$ and its unperturbed value $n_0$ \footnote{We assume a homogenous density along the $y$ direction, except when testing for dynamic instability.}.
A rigorous upper bound for the lowest-energy band can then be obtained by making use of the sum-rule result \cite{Pitaevskii16}
\begin{equation}
\epsilon(\mathbf{k}) \leq \hbar\sqrt{\frac{ m_1}{m_{-1}}} = \sqrt{\frac{ \epsilon_0(\mathbf{k})}{\chi(\mathbf{k})/2}} , \label{Eq:EUpBound}
\end{equation}
where $\epsilon_0(\mathbf{k})=\hbar^2k^2/2m$ is the noninteracting dispersion relation, and $m_p=\int d\omega \omega^pS(\mathbf{k},\omega)$ are the $p$-moments of the dynamic structure factor.  
Actually, the upper bound  (\ref{Eq:EUpBound})  provides a better estimate than the Feynman upper bound $\epsilon_F(\mathbf{k})=\hbar m_1(\mathbf{k})/m_0(\mathbf{k})= \epsilon_0(\mathbf{k})/S(\mathbf{k})$, where $S(\mathbf{k})=m_0$ is the static structure factor \cite{Pitaevskii16}. 
Furthermore, at finite temperature the knowledge of $\chi(\mathbf{k})$ provides important information on the density fluctuations, embodied by the static structure factor which obeys  the fluctuation dissipation theorem $S(\mathbf{k})\ge k_BT \chi(\mathbf{k})$ \cite{Pitaevskii16}.
This becomes an equality for weakly-interacting  gases when $k_BT \gg \epsilon(\mathbf{k})$, which should be readily accessible in current dipolar experiments where the maxon corresponds to a temperature $\sim 10$ nK \cite{Petter2018}.

\section{Roton-maxon dispersion}

As a realistic example we focus on a condensate of $^{164}$Dy atoms with a trapping frequency $\omega_z = 2\pi\times100$ Hz,  density of $n_0=$ 300  $\mu m^{-2}$, and a scattering length $a_s = 85.5 a_0$, giving $\epsilon_{\rm dd}\approx1.5$. Three-body loses are expected to be minimal since the unperturbed peak 3D density is only $6.6\times 10^{19}$ m$^{-3}$ and the scattering length is well within the range already realized in experiments \cite{FerrierBarbut2018,Kadau2016a}.

In Fig.~\ref{Fig:Dispersion} (a) we show the static density response function $\chi$ calculated  {by applying a static periodic perturbation with wave vector $k_{\rm L}$} and using Eq.~(\ref{Eq:StatResPos}). For the dipolar condensate [Fig.~\ref{Fig:Dispersion} (a1)],
a large response peak dominates, indicative of a rotonized dispersion relation, see also \cite{JonaLasinio2013b}.  A similar sharp peak is known to characterize the static response of superfluid $^4$He as a consequence of the roton excitations \cite{Dalfovo1992}.
In contrast, for the non-dipolar condensate [Fig.~\ref{Fig:Dispersion} (a2)], the response is two orders of magnitude lower and monotonically decreases.

Excitation energies calculated from BdG theory (solid lines) are displayed in Fig.~\ref{Fig:Dispersion} (b1) for the dipolar condensate, and in Fig.~\ref{Fig:Dispersion} (b2) for a non-dipolar one. For the dipolar case, a roton-maxon character is clearly visible in the lowest band.   The upper bound [plus symbols (\ref{Eq:EUpBound})], involving the static response $\chi$, provides a very accurate prediction for the lowest band of the dipolar gas, practically indistinguishable from the BdG solution.
Such a result is highly nontrivial since our 3D calculations inherently include the contributions from higher bands [see Fig.~1 (b\ref{Fig:Dispersion})].
In contrast, for superfluid $^4$He the Feynman upper bound overestimates the roton energy by a factor of two \cite{Boronat1995}.
Figure \ref{Fig:Dispersion} (b2) shows that for the non-dipolar condensate the upper bound exhibits good agreement with the lowest band of the exact BdG energy only for $k_\rho l_z \lesssim 1$.
The sum-rule upper bound's success in predicting the roton-maxon dispersion is partly thanks to the low roton energy -- since the static response function is directly related to the inverse energy weighted sum-rule -- and partly due to the lowest band experiencing the most-attractive interactions at moderate to large $k_\rho$.
As an interesting side point: for the nondipolar condensate [Fig.~\ref{Fig:Dispersion} (b2)], the lowest bands tend to become degenerate in a pairwise fashion at large momentum. This behavior arises as the excitations become more surface-like \cite{Dalfovo1997} and the two planar surfaces essentially uncouple.

Determining the static  response $\chi$ directly using (\ref{Eq:Delta_n}) will likely require  high-resolution {\it in situ} imaging, which is now available in dipolar experiments \cite{Kadau2016a}. 
As an alternative observable, it is also convenient to profit from the side-peaks of the momentum distribution (particle distribution function) arising at $\mathbf{k}_{\rho}=\mathbf{k}_{\pm{\rm L}}$ from the perturbation. The side-peaks are a consequence of Bose-Einstein condensation, which couples the density and particle response functions.
In the linear response and single mode approximations  \footnote{The single-mode approximation becomes an equality if only one excitation contributes to the density response at a given momentum.}\cite{Stringari2018}, the number of atoms in these side-peaks $N_{\pm\mathbf{k}_{\rm L}}$ relates to the dispersion as
\begin{equation}
\frac{N_{\pm\mathbf{k}_{\rm L}}}{N} = \frac{\chi^2(\mathbf{k}_{\rm L})V_{\rm L}^2}{4}  = \left(\frac{\epsilon_0(\mathbf{k}_{\rm L})V_{\rm L}}{\epsilon^2(\mathbf{k}_{\rm L})}\right)^2  ,\label{Eq:chiMtm}
\end{equation} 
where, in deriving the second equality, we have used the estimate (\ref{Eq:EUpBound})  for the excitation energy in terms of the static response.
  For sufficiently large values of $k_{\rm L}$ these peaks can be accurately measured in experiments via time-of-flight measurements.
The momentum space condensate wavefunction can be used to numerically calculate $N_{\pm\mathbf{k}_{\rm L}}$.
 We have checked, for the rotonized dipolar condensate, that the numerical predictions for $\epsilon(\mathbf{k})$ extracted from (\ref{Eq:chiMtm}) also agree well with the ones previously  calculated using BdG theory, thereby opening a complementary approach for the experimental determination of the roton-maxon excitation spectrum.

\section{Dynamics after lattice removal}

Another approach for extracting the roton-maxon spectrum is to suddenly remove the perturbing lattice, and then to follow the ensuing in-trap dynamics either with the position space observable $\Delta n(t)$ (\ref{Eq:Delta_n}) or with momentum space observable $N_{\mathbf{k}_{\rm L}}(t)$.
A clear experimental advantage of directly measuring the oscillation frequency is that the dispersion relation can be extracted without the need for precise calibration of the lattice strength nor the density response amplitude.
For reference, the roton minimum in Fig.~\ref{Fig:Dispersion} corresponds to a wavelength of 4.3$\mu$m, a value that should be reasonably well-resolved in the current generation of experiments with \emph{in situ} imaging resolution of around $1~\mu$m \cite{Kadau2016a}.

\begin{figure}
\begin{center}
\includegraphics[width=3.4in]{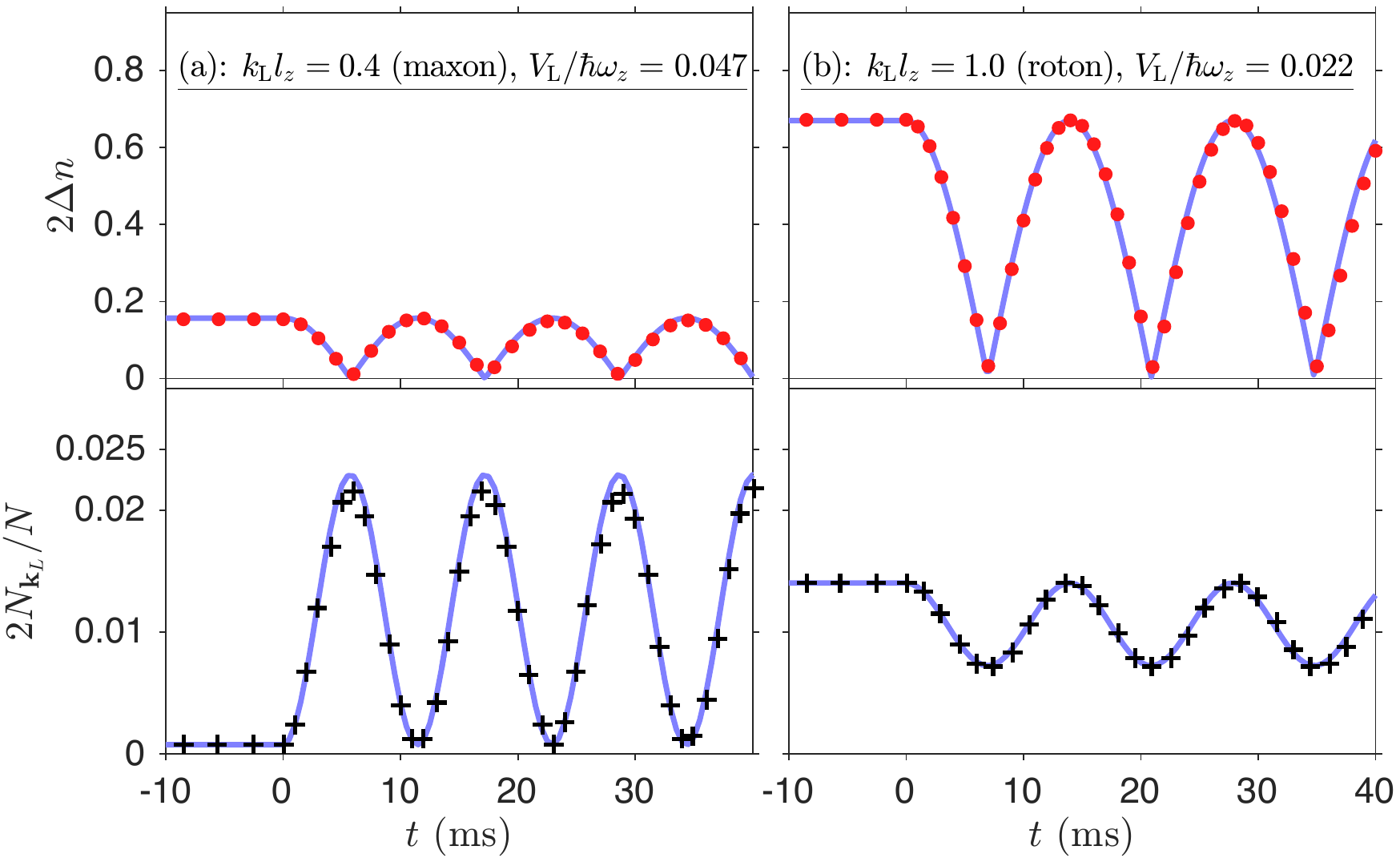}
\caption{ {In-trap dynamics of the density contrast $\Delta n$ [Eq.~(\ref{Eq:Delta_n})] and momentum response $N_{\mathbf{k}_{\rm L}}$ after sudden lattice removal at $t=0$. Lattices with (a) a maxon and (b) a roton wavelength are considered.
The single-mode predictions (\ref{Eq:Dnt}) and (\ref{Eq:NkL}) -- with $\epsilon$ extracted from Fig.~\ref{Fig:Dispersion} -- are shown as solid blue lines, while the time-dependent GPE results appear as symbols.}
\label{Fig:Dyn_NL}}
\end{center}
\end{figure}

\begin{figure}
\begin{center}
\includegraphics[width=3.4in]{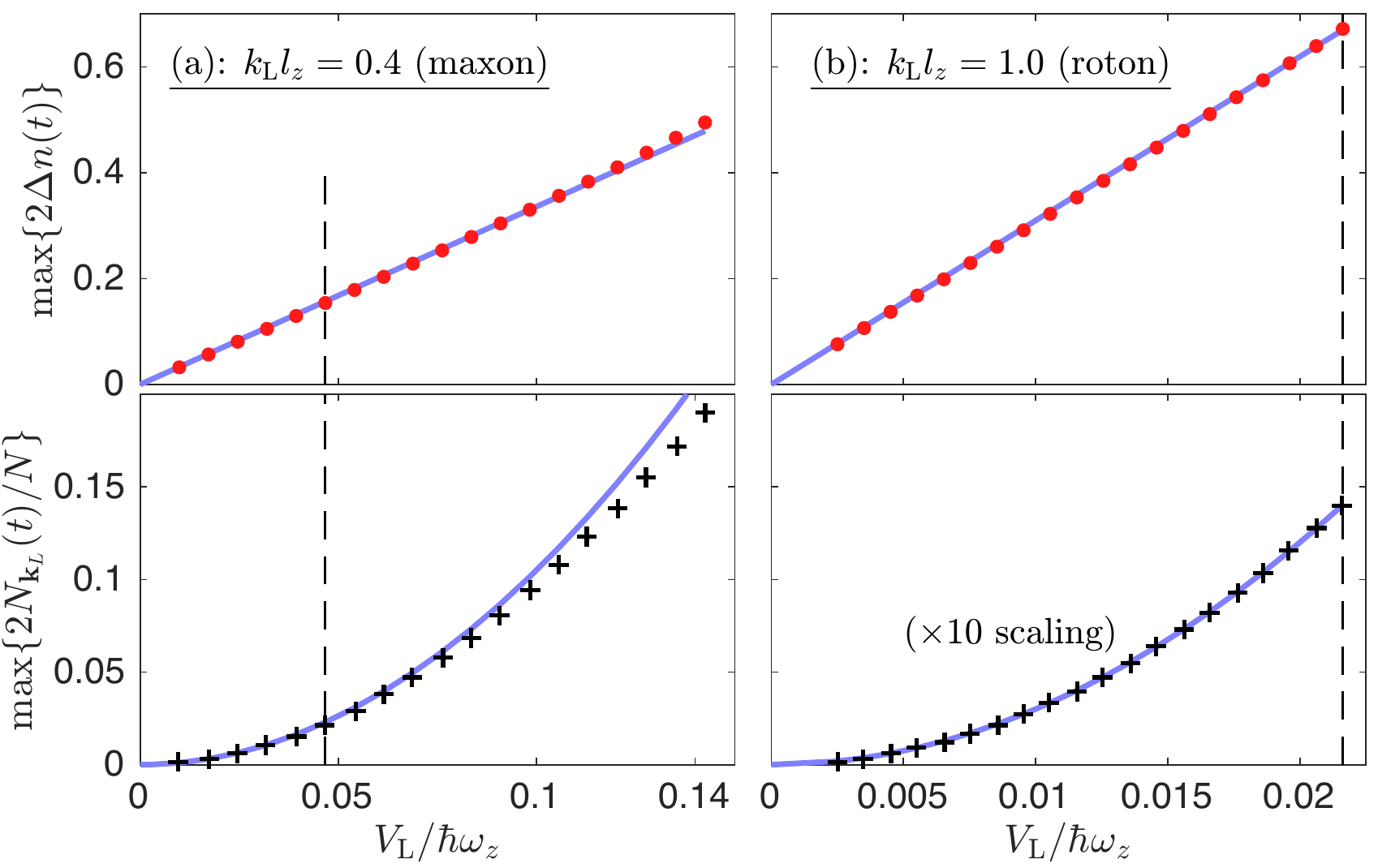}
\caption{ In-trap oscillation maxima of the density contrast and the momentum response (see Fig.~\ref{Fig:Dyn_NL}), as a function of the static lattice strength prior to its sudden removal.
As in Fig.~\ref{Fig:Dyn_NL}, here we consider (a) a maxon and (b) a roton.
The linear responses (\ref{Eq:Dnt}) and (\ref{Eq:NkL}) -- using BdG energies -- are shown as solid blue lines, while GPE results appear as symbols. The vertical dashed lines represent the $V_{\rm L}$ considered in Fig.~\ref{Fig:Dyn_NL}.
\label{Fig:E_V0}}
\end{center}
\end{figure}

We simulate this starting with a  ground state  
 in the presence of the lattice and then evolve it according to the GPE (\ref{Eq:GPE}) with the lattice suddenly removed (i.e.~$V_{\rm L}=0$ for $t>0$).
Such GPE dynamics are shown as symbols in Fig.~\ref{Fig:Dyn_NL} (a) for a lattice near the maxon wavelength ($k_{\rm L}l_z=0.4$), and in Fig.~\ref{Fig:Dyn_NL} (b) for a roton ($k_{\rm L}l_z=1$).
Both $\Delta n$ and  $N_{\mathbf{k}_{\rm L}}$ are seen to exhibit oscillations at twice the frequency of the dispersion relation.
{From an analytic perspective we can also predict these quantities using linear response theory, where in the single mode approximation they are:}
\begin{equation}
\Delta n(t) = \frac{4V_{\rm L} \epsilon_0(\mathbf{k}_{\rm L})}{\epsilon^2(\mathbf{k}_{\rm L})} \left|\cos\left(\frac{\epsilon(\mathbf{k}_{\rm L})t}{\hbar} \right)\right| , \label{Eq:Dnt}
\end{equation} 
\begin{align}
\!\!\!\!\frac{N_{\mathbf{k}_{\rm L}}(t)}{N}\! = \!\frac{V_{\rm L}^2}{\epsilon^2(\mathbf{k}_{\rm L})} \Bigg[ \frac{\epsilon_0^2(\mathbf{k}_{\rm L})}{\epsilon^2(\mathbf{k}_{\rm L})} \cos^2 \frac{\epsilon(\mathbf{k}_{\rm L})t}{\hbar} + \sin^2 \frac{\epsilon(\mathbf{k}_{\rm L})t}{\hbar}  \Bigg] . \label{Eq:NkL}
\end{align}
Equations (\ref{Eq:Dnt}) and (\ref{Eq:NkL}) are included in Fig.~\ref{Fig:Dyn_NL} as solid blue lines, where their excellent agreement with the symbols confirms that the GPE oscillation frequencies are indeed representative of the lowest-band dispersion [Fig.~\ref{Fig:Dispersion} (b1)].

While, as expected, Fig.~\ref{Fig:Dyn_NL} shows that $\Delta n(t)$ always decreases immediately after the lattice  is removed (at $t=0$), it is interesting to note that the behavior for $N_{\mathbf{k}_{\rm L}}(t)$ is qualitatively different.
Although $N_{\mathbf{k}_{\rm L}}(t)$ initially decreases for the roton case (b), it instead sharply increases for the maxon case (a). From a detectability viewpoint, these large upward oscillations for maxons should more than compensate for their relatively weak static response ($t<0$).
This behavior can be explained by considering (\ref{Eq:Dnt}) and (\ref{Eq:NkL}) in light of the effective interactions. For a non-interacting BEC one has $\epsilon=\epsilon_0$, which gives the intuitive result that  $\Delta n(t)$ oscillates while $N_{\mathbf{k}_{\rm L}}(t)$ remains constant.
Maxons (as well as phonons) have $\epsilon_0/\epsilon<1$ because of an effectively repulsive interaction at the relevant wavevector. At the moment that the lattice is removed the density perturbation is maximal and hence so too is the interaction energy. A quarter of an excitation period later, the density is flat and the interaction energy is now minimal, {with the difference being converted into} kinetic energy which manifests as an increase of $N_{\mathbf{k}_{\rm L}}$.
Rotons experience an effectively attractive interaction, hence  $\epsilon_0/\epsilon>1$, which explains why their oscillatory behavior is reversed \footnote{It should be noted that when we say that the interactions are effectively repulsive or attractive we are referring to the net contribution from the interactions to the BdG energies. Figure 5(a) of \cite{Blakie2013a} demonstrates how the effective interactions (denoted $\tilde{U}_{2D}$ -- also see appendix therein) change sign at around $k_\rho a_\rho \sim 3$, where $a_\rho = \sqrt{\hbar/m\omega_\rho}$, for radial trap frequency $\omega_\rho = 2\pi f_\rho$. For reference, the maxon wavelength there is at around $k_\rho a_\rho \sim 2$, and the roton wavelength is at $k_\rho a_\rho \sim 5$. Note that Fig.~5(a) of \cite{Blakie2013a} also demonstrates that the quantum depletion (the $T=0$ result) exhibits a `hole' where the effective interactions vanish at $k_\rho a_\rho \sim 3$, as expected.}.

\section{Extent of the linear regime}

Larger perturbations will be easier to detect, but may deviate from the linear response regime. Additionally, large perturbations can trigger the rotonized condensate to collapse \cite{Corson2013a,Corson2013b}.
In Fig.~\ref{Fig:E_V0}, we address these issues with the same two lattice wavelengths as in Fig.~\ref{Fig:Dyn_NL}, i.e.~(a) a maxon and (b) a roton.
{GPE results are shown as symbols and we see that $\max\{\Delta n(t)\}\propto V_{\rm L}$, while $\max\{N_{\mathbf{k}_{\rm L}}(t)\}\propto V_{\rm L}^2$, in good agreement with the predictions from Eqs.~(\ref{Eq:Dnt}) and (\ref{Eq:NkL}).
We have checked that for all $V_L$ considered, the in-trap oscillation frequencies (see Fig.~\ref{Fig:Dyn_NL}) coincide with high precision (within $1\%$) to the BdG roton-maxon frequencies in Fig.~\ref{Fig:Dispersion}. 
In fact, the excellent agreement in Fig.~\ref{Fig:Dyn_NL} (b) is for one of the most nonlinear cases, having a density contrast of $\max\{2\Delta n(t) \}\approx 0.7$, as indicated by the dashed line in Fig.~\ref{Fig:E_V0} (b).
This robustness of the linear response regime is important for the usefulness of our approaches.
Similarly, the GPE energy predictions extracted from the static density response $\Delta n(t=0)$ [using (7)] show excellent agreement with the BdG energies (within $1\%$), and the excitation energies inferred from $N_{\mathbf{k}_{\rm L}}(t=0)$ [using (8)] agree to within $0.025\hbar\omega_z$ for the regimes considered.

It should be noted that all results shown in Fig.~\ref{Fig:E_V0} are within the stable regime. For larger $V_{\rm L}$, the stationary states become dynamically unstable and the remaining translational symmetry breaks, i.e.~the high-density stripes break up to form quantum droplets \cite{Kadau2016a,Chomaz2016,Schmitt2016}. Despite this, the stability window should be large enough since $N_{\mathbf{k}_{\rm L}}$ is sizeable and {the density contrast is already quite large, i.e.~$2\Delta n\sim 0.5$.}

\section{Conclusions}

We have outlined novel approaches for the quantitative extraction of dispersion relations in quantum gases, focussing on the roton-maxon spectrum of dipolar BECs to demonstrate their effectiveness.
By measuring the static density response in position space -- or the corresponding side peaks of the momentum distribution -- a sum-rule upper bound provides an almost exact prediction for the roton-maxon dispersion of the lowest band, as well as the phonon spectrum for non-dipolar BECs. This is remarkable given that the Feynman sum-rule approach for superfluid $^4$He overestimates the roton energy by a factor of two.
By suddenly removing the lattice and observing the ensuing in-trap dynamics, we demonstrated that 
both the density and momentum side peaks
 oscillate in a stable manner at twice the characteristic frequency of the dispersion relation. Crucial for experimental observability, the oscillation frequency remains constant even for large perturbation amplitudes.
Interestingly, the side peak weights of the momentum distribution oscillate oppositely for rotons as compared to phonons and maxons, presenting a clear signature for the effectively attractive interactions experienced by the rotons. We quantitatively explained this behavior using perturbation theory.

\vspace{1cm}
\section*{Acknowledgments}
\vspace{-0.2cm}

We acknowledge useful discussions with Lauriane Chomaz, Franco Dalfovo and Francesca Ferlaino.
This work was supported by the QUIC grant of the Horizon 2020 FET program, the Provincia Autonoma di Trento, and the DFG/FWF (FOR 2247). RNB was supported by the European Union's Horizon 2020 research and innovation programme under the Marie Sk{\l}odowska-Curie grant agreement No. 793504 (DDQF). PBB was supported by the Marsden Fund of New Zealand.

\bibliographystyle{apsrev4-1}


%

\end{document}